\documentclass[preprint]{aastex61}

\usepackage{verbatim, graphicx,  ifthen, pbox}
\usepackage{eso-pic}

\newcommand{\asec}{\prime\prime}
\newcommand{\UZ}{2014~UZ$_{224}$}

\submitjournal{ApJ Letters, accepted}

\shorttitle{Discovery of a Large Scattered Disk Object at 92 AU}
\shortauthors{Gerdes et al.}

\begin{document}

\title{Discovery and Physical Characterization of a Large Scattered Disk Object at 92 AU}

\correspondingauthor{David Gerdes}
\email{gerdes@umich.edu}

\author[0000-0001-6942-2736]{D.~W.~Gerdes}
\affiliation{Department of Physics, University of Michigan, Ann Arbor, MI 48109, USA}
\affiliation{Department of Astronomy, University of Michigan, Ann Arbor, MI 48109, USA}
\author[0000-0003-2764-7093]{M.~Sako}
\affiliation{Department of Physics and Astronomy, University of Pennsylvania, Philadelphia, PA 19104, USA}
\author[0000-0002-6126-8487]{S.~Hamilton}
\affiliation{Department of Physics, University of Michigan, Ann Arbor, MI 48109, USA}
\author[0000-0002-0661-7517]{K.~Zhang}
\affiliation{Department of Astronomy, University of Michigan, Ann Arbor, MI 48109, USA}
\author[0000-0001-7721-6457]{T.~Khain}
\affiliation{Department of Physics, University of Michigan, Ann Arbor, MI 48109, USA}
\author[0000-0002-7733-4522]{J.~C.~Becker}
\affiliation{Department of Astronomy, University of Michigan, Ann Arbor, MI 48109, USA}
\author[0000-0002-0609-3987]{J.~Annis}
\affiliation{Fermi National Accelerator Laboratory, P. O. Box 500, Batavia, IL 60510, USA}
\author[0000-0003-0072-6736]{W.~Wester}
\affiliation{Fermi National Accelerator Laboratory, P. O. Box 500, Batavia, IL 60510, USA}
\author{G.~M.~Bernstein}
\affiliation{Department of Physics and Astronomy, University of Pennsylvania, Philadelphia, PA 19104, USA}
\author[0000-0002-4340-1299]{C.~Scheibner}
\affiliation{Department of Physics, University of Michigan, Ann Arbor, MI 48109, USA}
\affiliation{Department of Physics, St. Olaf Collage, Northfield, MN 55057, USA}
\author[0000-0002-8906-2835]{L.~Zullo}
\affiliation{Department of Physics, University of Michigan, Ann Arbor, MI 48109, USA}
\author[0000-0002-8167-1767]{F.~Adams}
\affiliation{Department of Physics, University of Michigan, Ann Arbor, MI 48109, USA}
\affiliation{Department of Astronomy, University of Michigan, Ann Arbor, MI 48109, USA}
\author[0000-0003-4179-6394]{E.~Bergin}
\affiliation{Department of Astronomy, University of Michigan, Ann Arbor, MI 48109, USA}
\author[0000-0002-7123-8943]{A.~R.~Walker}
\affiliation{Cerro Tololo Inter-American Observatory, National Optical Astronomy Observatory, Casilla 603, La Serena, Chile}
\author{J.~H.~Mueller}
\affiliation{Illinois Mathematics and Science Academy, 1500 Sullivan Rd., Aurora, IL 60506-1000, USA}
\affiliation{Fermi National Accelerator Laboratory, P. O. Box 500, Batavia, IL 60510, USA}
\author{T. M. C.~Abbott}
\affiliation{Cerro Tololo Inter-American Observatory, National Optical Astronomy Observatory, Casilla 603, La Serena, Chile}
\author{F.~B.~Abdalla}
\affiliation{Department of Physics \& Astronomy, University College London, Gower Street, London, WC1E 6BT, UK}
\affiliation{Department of Physics and Electronics, Rhodes University, PO Box 94, Grahamstown, 6140, South Africa}
\author{S.~Allam}
\affiliation{Fermi National Accelerator Laboratory, P. O. Box 500, Batavia, IL 60510, USA}
\author{K.~Bechtol}
\affiliation{LSST, 933 North Cherry Avenue, Tucson, AZ 85721, USA}
\author{A.~Benoit-L{\'e}vy}
\affiliation{CNRS, UMR 7095, Institut d'Astrophysique de Paris, F-75014, Paris, France}
\affiliation{Department of Physics \& Astronomy, University College London, Gower Street, London, WC1E 6BT, UK}
\affiliation{Sorbonne Universit\'es, UPMC Univ Paris 06, UMR 7095, Institut d'Astrophysique de Paris, F-75014, Paris, France}
\author{E.~Bertin}
\affiliation{CNRS, UMR 7095, Institut d'Astrophysique de Paris, F-75014, Paris, France}
\affiliation{Sorbonne Universit\'es, UPMC Univ Paris 06, UMR 7095, Institut d'Astrophysique de Paris, F-75014, Paris, France}
\author{D.~Brooks}
\affiliation{Department of Physics \& Astronomy, University College London, Gower Street, London, WC1E 6BT, UK}
\author{D.~L.~Burke}
\affiliation{Kavli Institute for Particle Astrophysics \& Cosmology, P. O. Box 2450, Stanford University, Stanford, CA 94305, USA}
\affiliation{SLAC National Accelerator Laboratory, Menlo Park, CA 94025, USA}
\author{A. Carnero Rosell}
\affiliation{Laborat\'orio Interinstitucional de e-Astronomia - LIneA, Rua Gal. Jos\'e Cristino 77, Rio de Janeiro, RJ - 20921-400, Brazil}
\affiliation{Observat\'orio Nacional, Rua Gal. Jos\'e Cristino 77, Rio de Janeiro, RJ - 20921-400, Brazil}
\author{M.~Carrasco~Kind}
\affiliation{Department of Astronomy, University of Illinois, 1002 W. Green Street, Urbana, IL 61801, USA}
\affiliation{National Center for Supercomputing Applications, 1205 West Clark St., Urbana, IL 61801, USA}
\author{J.~Carretero}
\affiliation{Institut de F\'{\i}sica d'Altes Energies (IFAE), The Barcelona Institute of Science and Technology, Campus UAB, 08193 Bellaterra (Barcelona) Spain}
\author{C.~E.~Cunha}
\affiliation{Kavli Institute for Particle Astrophysics \& Cosmology, P. O. Box 2450, Stanford University, Stanford, CA 94305, USA}
\author{L.~N.~da Costa}
\affiliation{Laborat\'orio Interinstitucional de e-Astronomia - LIneA, Rua Gal. Jos\'e Cristino 77, Rio de Janeiro, RJ - 20921-400, Brazil}
\affiliation{Observat\'orio Nacional, Rua Gal. Jos\'e Cristino 77, Rio de Janeiro, RJ - 20921-400, Brazil}
\author{S.~Desai}
\affiliation{Department of Physics, IIT Hyderabad, Kandi, Telangana 502285, India}
\author{H.~T.~Diehl}
\affiliation{Fermi National Accelerator Laboratory, P. O. Box 500, Batavia, IL 60510, USA}
\author{T.~F.~Eifler}
\affiliation{Jet Propulsion Laboratory, California Institute of Technology, 4800 Oak Grove Dr., Pasadena, CA 91109, USA}
\author{B.~Flaugher}
\affiliation{Fermi National Accelerator Laboratory, P. O. Box 500, Batavia, IL 60510, USA}
\author{J.~Frieman}
\affiliation{Fermi National Accelerator Laboratory, P. O. Box 500, Batavia, IL 60510, USA}
\affiliation{Kavli Institute for Cosmological Physics, University of Chicago, Chicago, IL 60637, USA}
\author{J.~Garc\'ia-Bellido}
\affiliation{Instituto de Fisica Teorica UAM/CSIC, Universidad Autonoma de Madrid, 28049 Madrid, Spain}
\author{E.~Gaztanaga}
\affiliation{Institut de Ci\`encies de l'Espai, IEEC-CSIC, Campus UAB, Carrer de Can Magrans, s/n,  08193 Bellaterra, Barcelona, Spain}
\author{D.~A.~Goldstein}
\affiliation{Department of Astronomy, University of California, Berkeley,  501 Campbell Hall, Berkeley, CA 94720, USA}
\affiliation{Lawrence Berkeley National Laboratory, 1 Cyclotron Road, Berkeley, CA 94720, USA}
\author{D.~Gruen}
\affiliation{Kavli Institute for Particle Astrophysics \& Cosmology, P. O. Box 2450, Stanford University, Stanford, CA 94305, USA}
\affiliation{SLAC National Accelerator Laboratory, Menlo Park, CA 94025, USA}
\author{J.~Gschwend}
\affiliation{Laborat\'orio Interinstitucional de e-Astronomia - LIneA, Rua Gal. Jos\'e Cristino 77, Rio de Janeiro, RJ - 20921-400, Brazil}
\affiliation{Observat\'orio Nacional, Rua Gal. Jos\'e Cristino 77, Rio de Janeiro, RJ - 20921-400, Brazil}
\author{G.~Gutierrez}
\affiliation{Fermi National Accelerator Laboratory, P. O. Box 500, Batavia, IL 60510, USA}
\author{K.~Honscheid}
\affiliation{Center for Cosmology and Astro-Particle Physics, The Ohio State University, Columbus, OH 43210, USA}
\affiliation{Department of Physics, The Ohio State University, Columbus, OH 43210, USA}
\author{D.~J.~James}
\affiliation{Astronomy Department, University of Washington, Box 351580, Seattle, WA 98195, USA}
\affiliation{Cerro Tololo Inter-American Observatory, National Optical Astronomy Observatory, Casilla 603, La Serena, Chile}
\author{S.~Kent}
\affiliation{Fermi National Accelerator Laboratory, P. O. Box 500, Batavia, IL 60510, USA}
\affiliation{Kavli Institute for Cosmological Physics, University of Chicago, Chicago, IL 60637, USA}
\author{E.~Krause}
\affiliation{Kavli Institute for Particle Astrophysics \& Cosmology, P. O. Box 2450, Stanford University, Stanford, CA 94305, USA}
\author{K.~Kuehn}
\affiliation{Australian Astronomical Observatory, North Ryde, NSW 2113, Australia}
\author{N.~Kuropatkin}
\affiliation{Fermi National Accelerator Laboratory, P. O. Box 500, Batavia, IL 60510, USA}
\author{O.~Lahav}
\affiliation{Department of Physics \& Astronomy, University College London, Gower Street, London, WC1E 6BT, UK}
\author{T.~S.~Li}
\affiliation{Fermi National Accelerator Laboratory, P. O. Box 500, Batavia, IL 60510, USA}
\affiliation{George P. and Cynthia Woods Mitchell Institute for Fundamental Physics and Astronomy, and Department of Physics and Astronomy, Texas A\&M University, College Station, TX 77843,  USA}
\author{M.~A.~G.~Maia}
\affiliation{Laborat\'orio Interinstitucional de e-Astronomia - LIneA, Rua Gal. Jos\'e Cristino 77, Rio de Janeiro, RJ - 20921-400, Brazil}
\affiliation{Observat\'orio Nacional, Rua Gal. Jos\'e Cristino 77, Rio de Janeiro, RJ - 20921-400, Brazil}
\author{M.~March}
\affiliation{Department of Physics and Astronomy, University of Pennsylvania, Philadelphia, PA 19104, USA}
\author{J.~L.~Marshall}
\affiliation{George P. and Cynthia Woods Mitchell Institute for Fundamental Physics and Astronomy, and Department of Physics and Astronomy, Texas A\&M University, College Station, TX 77843,  USA}
\author{P.~Martini}
\affiliation{Center for Cosmology and Astro-Particle Physics, The Ohio State University, Columbus, OH 43210, USA}
\affiliation{Department of Astronomy, The Ohio State University, Columbus, OH 43210, USA}
\author{F.~Menanteau}
\affiliation{Department of Astronomy, University of Illinois, 1002 W. Green Street, Urbana, IL 61801, USA}
\affiliation{National Center for Supercomputing Applications, 1205 West Clark St., Urbana, IL 61801, USA}
\author{R.~Miquel}
\affiliation{Instituci\'o Catalana de Recerca i Estudis Avan\c{c}ats, E-08010 Barcelona, Spain}
\affiliation{Institut de F\'{\i}sica d'Altes Energies (IFAE), The Barcelona Institute of Science and Technology, Campus UAB, 08193 Bellaterra (Barcelona) Spain}
\author{R.~C.~Nichol}
\affiliation{Institute of Cosmology \& Gravitation, University of Portsmouth, Portsmouth, PO1 3FX, UK}
\author{A.~A.~Plazas}
\affiliation{Jet Propulsion Laboratory, California Institute of Technology, 4800 Oak Grove Dr., Pasadena, CA 91109, USA}
\author{A.~K.~Romer}
\affiliation{Department of Physics and Astronomy, Pevensey Building, University of Sussex, Brighton, BN1 9QH, UK}
\author{A.~Roodman}
\affiliation{Kavli Institute for Particle Astrophysics \& Cosmology, P. O. Box 2450, Stanford University, Stanford, CA 94305, USA}
\affiliation{SLAC National Accelerator Laboratory, Menlo Park, CA 94025, USA}
\author{E.~Sanchez}
\affiliation{Centro de Investigaciones Energ\'eticas, Medioambientales y Tecnol\'ogicas (CIEMAT), Madrid, Spain}
\author{I.~Sevilla-Noarbe}
\affiliation{Centro de Investigaciones Energ\'eticas, Medioambientales y Tecnol\'ogicas (CIEMAT), Madrid, Spain}
\author{M.~Smith}
\affiliation{School of Physics and Astronomy, University of Southampton,  Southampton, SO17 1BJ, UK}
\author{R.~C.~Smith}
\affiliation{Cerro Tololo Inter-American Observatory, National Optical Astronomy Observatory, Casilla 603, La Serena, Chile}
\author{M. Soares-Santos}
\affiliation{Fermi National Accelerator Laboratory, P. O. Box 500, Batavia, IL 60510, USA}
\author{F.~Sobreira}
\affiliation{Laborat\'orio Interinstitucional de e-Astronomia - LIneA, Rua Gal. Jos\'e Cristino 77, Rio de Janeiro, RJ - 20921-400, Brazil}
\affiliation{Universidade Federal do ABC, Centro de Ci\^encias Naturais e Humanas, Av. dos Estados, 5001, Santo Andr\'e, SP, Brazil, 09210-580}
\author{E.~Suchyta}
\affiliation{Computer Science and Mathematics Division, Oak Ridge National Laboratory, Oak Ridge, TN 37831}
\author{M.~E.~C.~Swanson}
\affiliation{National Center for Supercomputing Applications, 1205 West Clark St., Urbana, IL 61801, USA}
\author{G.~Tarle}
\affiliation{Department of Physics, University of Michigan, Ann Arbor, MI 48109, USA}
\author{D.~L.~Tucker}
\affiliation{Fermi National Accelerator Laboratory, P. O. Box 500, Batavia, IL 60510, USA}
\author{Y.~Zhang}
\affiliation{Fermi National Accelerator Laboratory, P. O. Box 500, Batavia, IL 60510, USA}

\collaboration{(DES Collaboration)}

\AddToShipoutPictureBG*{%
  \AtPageUpperLeft{%
    \hspace{0.7\paperwidth}%
    \raisebox{-10\baselineskip}{%
      \makebox[0pt][l]{\textnormal{DES 2016-0198}}   % DES publication number
}}}%

\AddToShipoutPictureBG*{%
  \AtPageUpperLeft{%
    \hspace{0.7\paperwidth}%
    \raisebox{-11\baselineskip}{%
      \makebox[0pt][l]{\textnormal{Fermilab PUB-17-027-AE}}   %Fermilab preprint number
}}}%

\begin{abstract}
We report the observation and physical characterization of the possible dwarf planet \UZ\ (``DeeDee''), a dynamically detached trans-Neptunian object discovered at 92 AU.  This object is currently the second-most distant 
known trans-Neptunian object with reported orbital elements, surpassed in distance only by the dwarf planet Eris. The object was discovered with an $r$-band magnitude of 23.0 in data collected by the Dark Energy Survey between 2014 and 2016.  Its 1140-year orbit has  $(a,e,i) = (109~\mathrm{AU}, 0.65, 26.8\arcdeg)$. It will reach its perihelion distance of 38 AU in the year 2142.
Integrations of its orbit show it to be dynamically stable on Gyr timescales, with only weak interactions with Neptune. We have performed 
followup observations with ALMA, using 3 hours of on-source integration time to measure the object's thermal emission in the Rayleigh-Jeans tail. The signal is detected at 
7$\sigma$ significance, from which we determine a $V$-band albedo of $13.1^{+3.3}_{-2.4}\mathrm{(stat)}^{+2.0}_{-1.4}\mathrm{(sys)}$ percent and a diameter of $635^{+57}_{-61}\mathrm{(stat)}^{+32}_{-39}\mathrm{(sys)}$~km, assuming a spherical body with uniform 
surface properties.
\end{abstract}

\keywords{Kuiper belt: general --- infrared: planetary systems --- methods: observational --- techniques: photometric}

\section{Introduction}

The scattered disk and inner Oort cloud populations of trans-Neptunian objects (TNOs) extend well beyond the classical Kuiper Belt, to distances of hundreds of AU. 
%Nine of the ten largest TNOs belong to these populations, with a median inclination of 21$^{\circ}$ \citep{Sheppard2011}. 
These dynamically disturbed populations must have arisen from very different mechanisms
than those that produced the classical Kuiper Belt, as evidenced by marked differences in their sizes \citep{Fraser2014}, colors \citep{Tegler2000}, albedos \citep{Brucker2009}, and fraction of binaries \citep{Noll2008}.
The scattered disk population has been further divided by \citet{Gladman2008} into objects which are actively scattering off Neptune (as indicated by a significant variation in their semi-major axis on 10~Myr timescales), and detached objects (non-scattering, non-resonant objects with $e>0.24$). The half-dozen longest-period members of these populations display a statistically improbable clustering in argument of perihelion and longitude of ascending node. This finding
 has motivated the hypothesis of a distant
super-earth \citep{Trujillo2014, BatyginBrown2016}, sometimes called Planet 9. Deep, wide-area surveys capable of probing the distant scattered disk to high ecliptic latitudes have considerable potential to contribute to our knowledge of this region \citep{Abbott2016}. In this Letter we report the discovery of a large scattered disk object at 92 AU using data from the Dark Energy Survey  (DES; \citealt{Flaugher2005}), with followup radiometric measurements by ALMA.  Of known Solar System objects with reported orbital elements, only the Pluto-sized dwarf planet Eris is currently more distant.  

The DES is an optical survey of 5000 square degrees of the southern sky being carried out with the Dark Energy Camera (DECam, \citealt{Flaugher2015}) on the 4-meter Blanco telescope at Cerro Tololo Inter-American Observatory in Chile. DECam is a prime-focus camera with a 3 square degree field of view and a focal plane consisting of 62 2k$\times$4k fully-depleted, red-sensitive CCDs. To achieve its primary scientific goal of constraining the dark energy equation of state, the DES has been awarded 525 nights over 5 years to carry out two interleaved surveys. The DES Supernova Program (DES-SN, \citealt{Bernstein2012}) 
images ten distinct DECam fields (a total of 30 sq.\ deg.) in the $griz$ bands at approximately weekly intervals throughout the DES observing season, which runs from mid-August through mid-February. The Wide Survey covers the full survey footprint in the $grizY$ bands to a limiting single-exposure depth of $m_{r}\sim 23.8$, with the goal of achieving 10 tilings per filter over the duration of the survey. The same combination of survey area and depth
that makes DES a powerful tool for precision cosmology also makes it well suited to identify faint, distant objects in our own Solar System. With broad off-ecliptic coverage, it is especially well-suited to identifying members of the 
scattered disk and other high-inclination TNO populations such as detached and inner Oort cloud objects. We have previously reported on searches for TNOs in the DES-SN fields from the 
first two DES seasons, where discoveries have included two Neptune Trojans \citep{Gerdes2016} and the ``extreme TNO'' 2013~RF$_{98}$ \citep{Abbott2016} whose orbital alignment with other members of its class helped motivate the Planet 9 hypothesis. This paper presents our first result from the extension of the TNO search to the full DES Wide Survey, using data collected during the first three DES observing campaigns between August 2013 and February 2016. 

\section{Optical Data and Analysis}

%\begin{figure*}[htbp]
%\begin{center}
%\includegraphics[width=0.4\linewidth]{year-1-nepochs} \\
%\includegraphics[width=0.4\linewidth]{year-2-nepochs} \\
%\includegraphics[width=0.4\linewidth]{year-3-nepochs}
%\caption{\label{fig:coverage-maps} Coverage map of the DES Wide Survey for the first three observing campaigns. The color scale shows the number of nights in which each region of the survey
%was targeted in at least one $g$, $r$, $i$, or $z$-band exposure during the given season. Observations on a minimum of three different nights at a given opposition are required to form a TNO candidate.}
%\end{center}
%\end{figure*}

This analysis uses data from 14,857 exposures collected in the $griz$ bands during the first 3 DES observing campaigns \citep{Diehl2016}. 
These exposures cover a 2500 square degree region north of DEC$=-40$, about half the full survey area. They contain over 1.1 billion individual object detections. 

We identify transient objects using a variant of
  the DES supernova difference-imaging pipeline, \texttt{DiffImg} \citep{Kessler2015}.  Each
exposure (search image) is subtracted from every other DES exposure
  (template image) of that region taken in the same band.  We do not use template images from
  the same night to avoid subtracting out the most distant and slowest moving
  objects, which may appear stationary over a period of several hours.

  The difference images created from each search-template pair are then
  averaged, and statistically significant sources are identified in the
  combined image.  Subtraction artifacts are rejected using a machine-learning
  technique described in \citet{Goldstein2015}.  This typically yields $\sim
  10$ good-quality transient detections on each $9\arcmin \times 18\arcmin$ area covered by a single
  CCD.

%  We do not search in the $Y$-band images because of their significantly lower depth compared to the other bands.

 After removal of stationary objects and artifacts with \texttt{DiffImg}, our search sample contains about 5 million single-epoch transients. While our selection efficiently retains true astrophysical transients---asteroids, variable stars, supernovae, etc.---the fraction of TNOs in this sample is on the order of only 0.1\%.

The apparent motion of a distant Solar System object over periods of several weeks is primarily due not to its own orbital motion but to parallax arising from the motion of the Earth. Our TNO search procedure begins by 
identifying pairs of detections within 30 nights of each other whose separation is consistent with the seasonally-appropriate parallax expected for a distant object ($\lesssim 4^{\asec}$/hour). We then attempt to link these pairs into chains of three or more observations, testing each chain for goodness of fit to an orbit using code built on the \texttt{fit\_radec} algorithm of \citet{Bernstein2000} (B\&K) and requiring $\chi^{2}/N<2.$

\begin{figure}[htbp]
\begin{center}
\includegraphics[width=\linewidth]{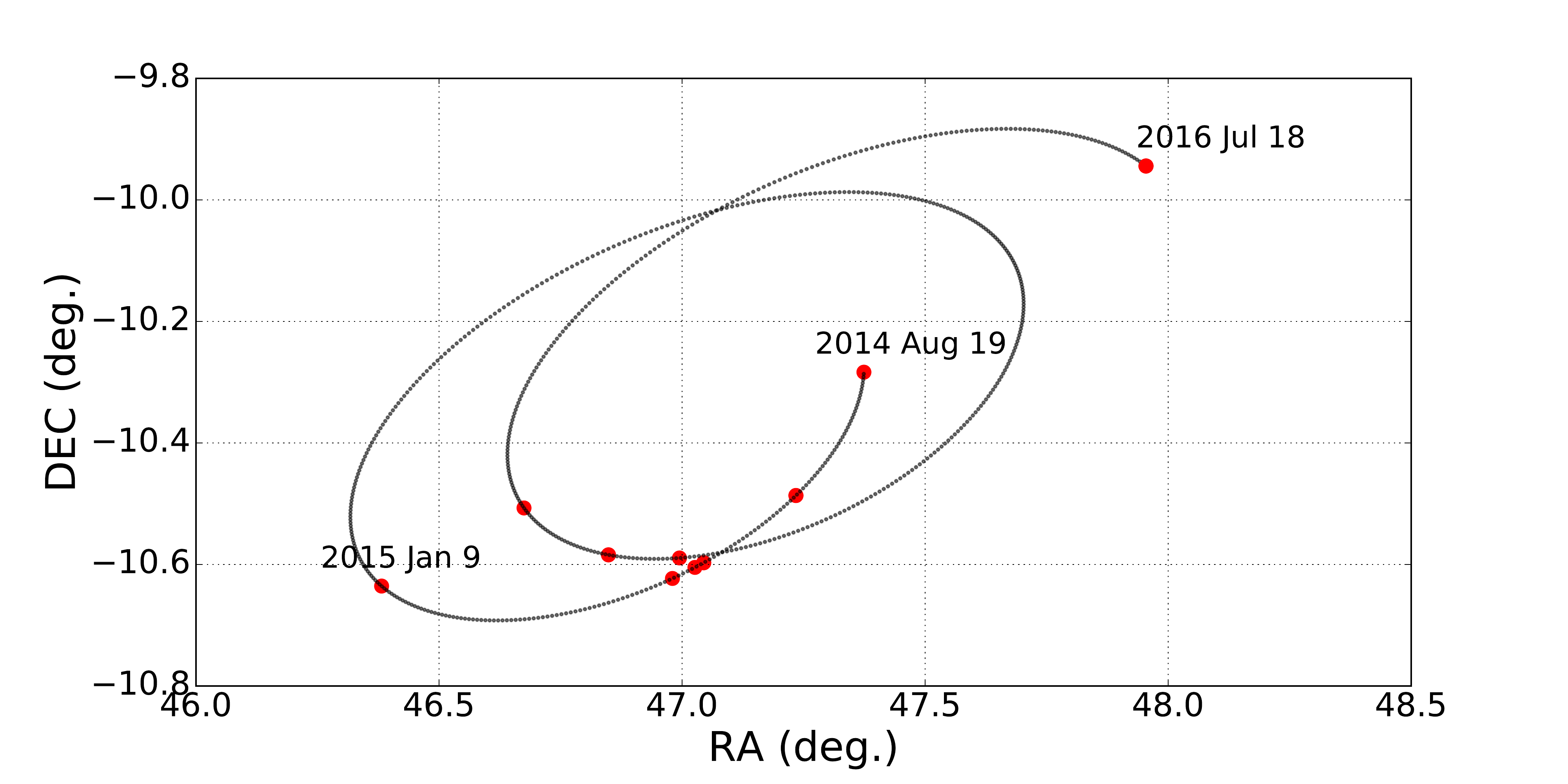}
\caption{\label{fig:trajectory} The path of \UZ\ over the course of its observed 699-day arc. Dots indicate locations at which the object was observed by the DES.}
\end{center}
\end{figure}

\UZ\ was originally detected at a heliocentric distance of 92.5~AU in 7 linked observations on 4 nights between 2014/9/27 and 2014/10/28, with an $r$-band magnitude of 23.0 and an ecliptic latitude of $-10.3^{\circ}$. 
The object was detected in 6 more DES survey images between 2014/8/19 and 2015/1/8, and was recovered in a targeted DECam observation on 2016/7/18. The motion of the object over the period of these observations is shown in Figure~\ref{fig:trajectory}. The orbital elements are obtained using the B\&K fitter. These and other data from these observations are shown in Table~\ref{tab:parameters}. We refer informally to this object as ``DeeDee,'' for ``distant dwarf.''

\begin{deluxetable}{ll}
\tabletypesize{\scriptsize}
\tablecaption{Orbital elements and other properties of \UZ. \label{tab:parameters} }
\tablewidth{0pt}
\tablehead{
\colhead{Parameter} & Value 
}
\startdata
$a$~(AU) & $108.90 \pm 7.36$ \\
$e$ & $0.651 \pm 0.030$ \\
$i$ (deg) & $26.78509 \pm0.00012$   \\
$\omega$ (deg) & $29.55 \pm 1.46$   \\
$\Omega$ (deg) & $131.142 \pm  0.053$ \\
Perihelion (AU) & $37.97 \pm 0.69$ \\
Perihelion date & 2142/01/02 $\pm$ 1654d  \\
Aphelion (AU) & $179.8 \pm 12.1$ \\
Period (yr) & $1136 \pm 115$ \\
Epoch JD & 2457600.5 \\
Heliocentric distance at discovery (AU) & 92.5 \\
Arc length (days) & 699 \\
Apparent mag ($r$) & $22.98 \pm 0.04$ \\
Apparent mag ($V$) & $23.38 \pm 0.05$ \\
Absolute mag $H_{V}$ & 3.5 \\
$g-r$ (mag.) & $0.77 \pm 0.11$ \\
$r-i$ (mag.) & $0.39 \pm 0.07$ \\
$i-z$ (mag.) & $  0.22 \pm 0.16 $ \\
Albedo (\%) & $13.1^{+3.3}_{-2.4}\mathrm{(stat)}^{+2.0}_{-1.4}\mathrm{(sys)}$ \\
Diameter (km) & $635^{+57}_{-61}\mathrm{(stat)}^{+32}_{-39}\mathrm{(sys)}$  \\
\enddata
%% Text for table notes should follow after the \enddata but before
%% the \end{deluxetable}. Make sure there is at least one \tablenotemark
%% in the table for each \tablenotetext.
\end{deluxetable}
Apparent and absolute magnitudes of Solar System objects are often standardized to Johnson-Cousins $V$-band magnitudes. We first derived transformation equations for stellar psf magnitudes to relate DES and SDSS magnitudes, then applied the transformations of \citet{Smith2002} to convert from the SDSS to Johnson-Cousins systems, obtaining $m_{V}=23.38\pm 0.05$. The transformation equations depend on the $g-r$ color of the object in question, which is uncertain at the level of $0.11$~mag. As a cross-check, the measured spectra of five TNOs with similar colors were flux-corrected and found to have a reasonable match to the observed DES magnitudes. From the flux-calibrated spectra of each of these TNOs, we applied a synthetic determination of the $V$-band magnitude. The central value and spread of these values is consistent with our measurement.

\section{Orbital Dynamics}

We next investigated the dynamical behavior of \UZ\ on Gyr timescales. We generated 100,000 clones of \UZ\ with respect to the best-fit orbit and its covariance matrix as described in \citet{Gladman2008}. Taking the clone with the smallest RMS residual to be the new best-fit, we repeated the clone-generating procedure and identified the clones that yield residuals consistent with observations. Out of these objects, we chose the clones with minimum and maximum semi-major axes, as well as five additional clones interspersed between those two, and numerically integrated the Solar System using all eight clones as test particles. We ran the integration for 1 Gyr using the hybrid symplectic and Bulirsch-Stoer integrator built into \texttt{Mercury6} \citep{Chambers1999}, and conserved energy to 1 part in $10^{9}$.
We did not include the terrestrial planets in our integrations, and we replaced Jupiter, Saturn, and Uranus with a solar $J_{2}$ \citep[as done in][]{BatyginBrown2016}. We included Neptune as an active body, because \UZ's perihelion distance of 38~AU brings it into proximity with Neptune. 

\begin{figure}[t!]
\epsscale{1}
  \begin{center}
      \leavevmode
\plotone{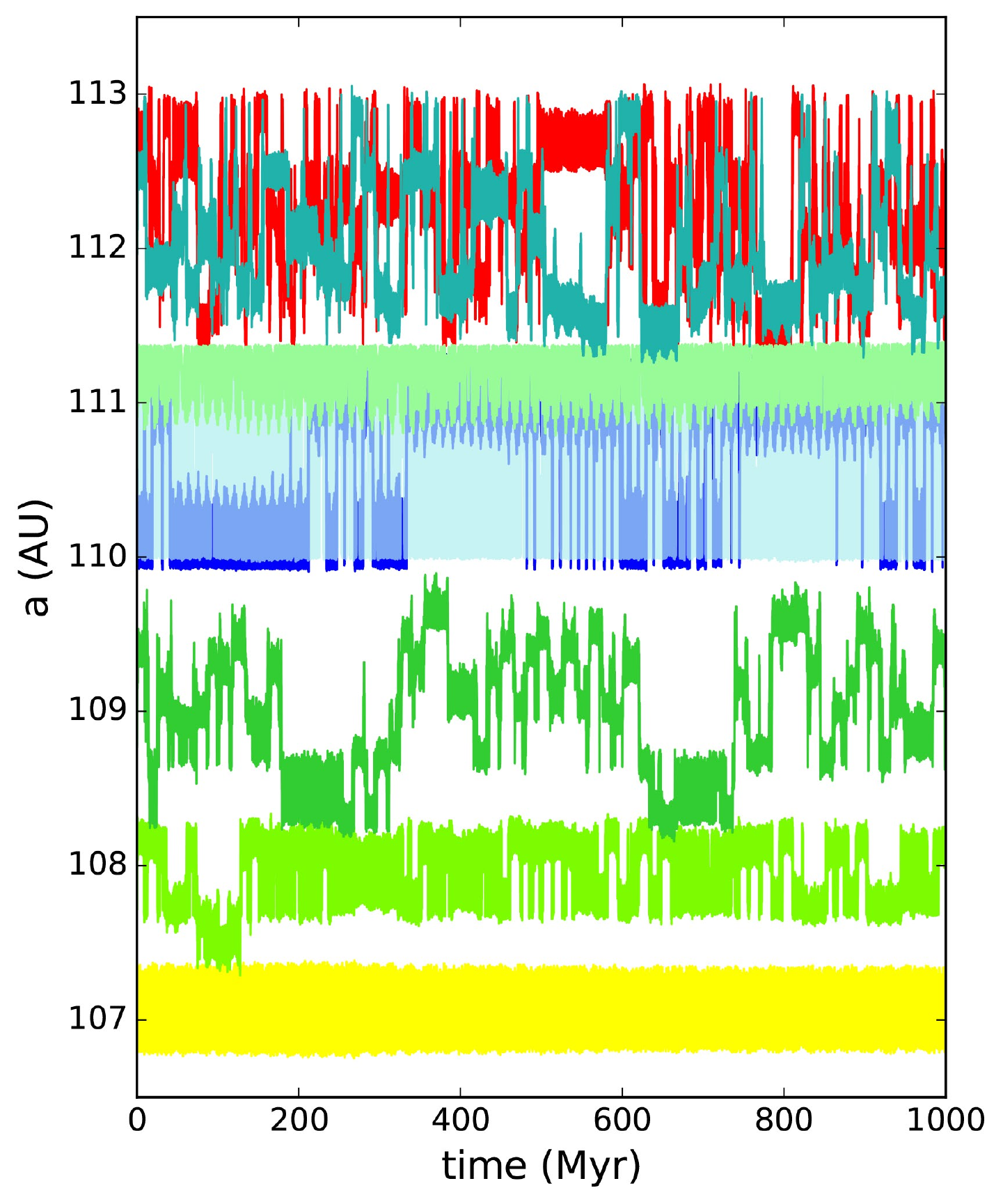}
\caption{\label{fig:dyn}Time evolution of semi-major axis over 1 Gyr for each of the eight clones of \UZ\ considered in this work, from the minimum (bottom line) to maximum (top) initial semi-major axis. For all clones,  $\delta a/a$ is less than 1\% in amplitude, demonstrating the long-term dynamical stability of this object in the presence of Neptune. } 

\end{center}
\end{figure}

As shown in Figure \ref{fig:dyn}, over 1 Gyr timescales each clone remains confined to a region closely surrounding its measured orbit, with $\delta a/a$ being less than 1\% for all clones. This result indicates that despite the potentially destabilizing interactions with Neptune, this object remains dynamically stable and satisfies the formal criteria of \citet{Gladman2008} as a detached TNO. Although the uncertainty on the object's semi-major axis overlaps with the 7:1 mean-motion resonance with Neptune, none of the clones we examined undergoes libration. We also performed several 4.5 Gyr integrations of the best-fit orbit. The object demonstrated stability over the full Solar System lifetime as well.  

\section{Measurement of Thermal Emission, Size, and Albedo}

We observed \UZ\ with director's discretionary time on the Atacama Large Millimeter/submillimeter Array (ALMA) on 2016 August 19 and 20. The observations were carried out with 41 antennae and baselines between 15-1462 meters. The source was tracked using a user-provided ephemeris. The correlator was configured to observe four continuum spectral windows centered on 224, 226, 240, and 242 GHz, respectively, resulting in a total bandwidth of 7.5GHz. The nearby quasars J0522-3627 and J0238+1636 were used as bandpass calibrators for the first night and the second observations, respectively. The amplitude and phase of observations were calibrated by J0257-1212, and J0423-0120 was used for absolute flux calibration. The total on-source integration time was 176 minutes. 

\begin{figure}[htbp]
\begin{center}
\includegraphics[width=\columnwidth]{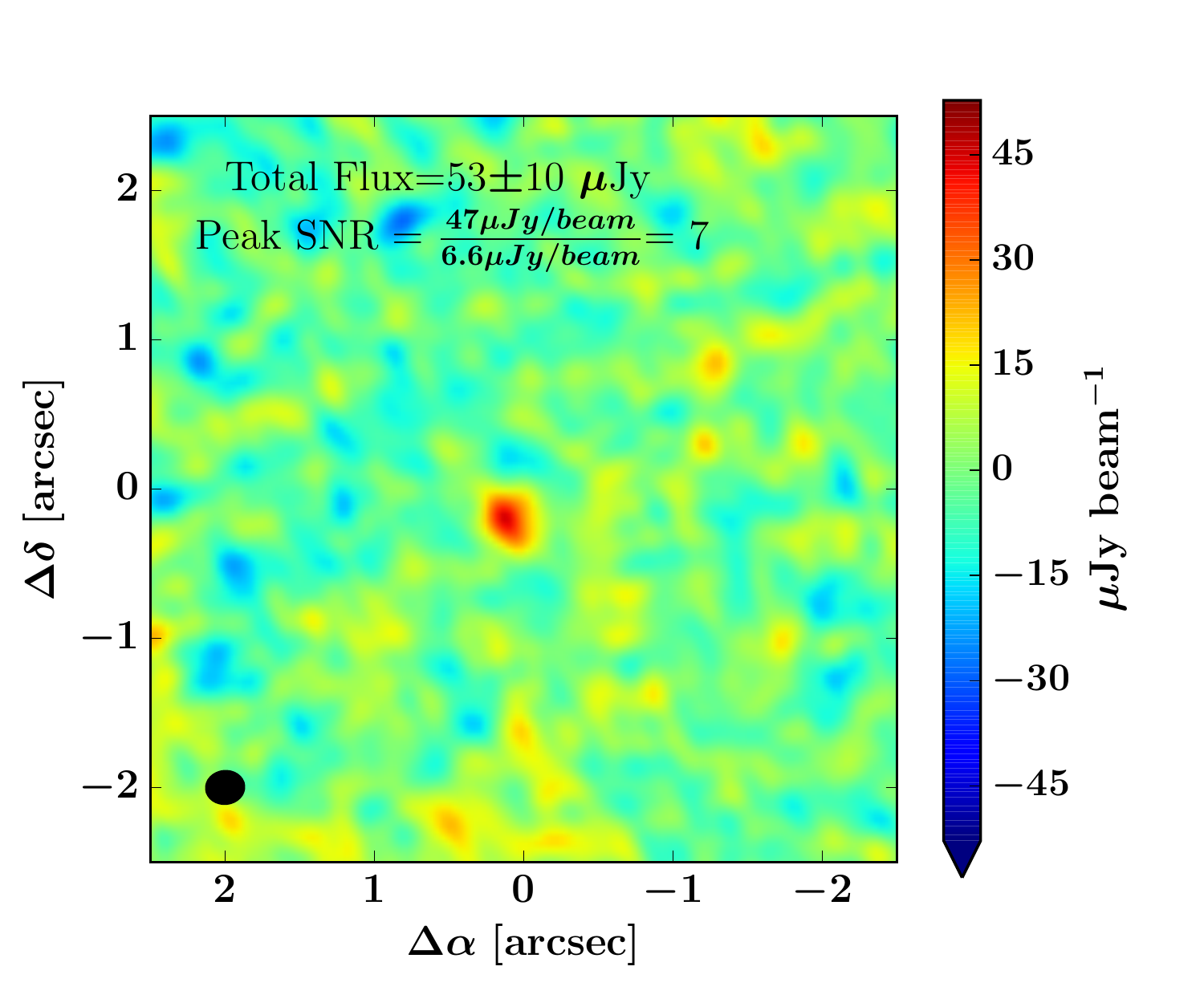}
\caption{\label{fig:ALMA-image} Calibrated, stacked image of \UZ\ from 3 hours of on-source integration with ALMA. The black ellipse represents the size of the synthesized beam.}
\end{center}
\end{figure}

The raw data were calibrated by NRAO staff manually using the CASA package version 4.6. The calibrated visibilities of five data sets were then stacked to align the position using the  \texttt{fixplanet} command. We generated a synthesized continuum image with the CLEAN algorithm and a natural weighting in CASA. The resulting synthesized beam is $0.30^{\asec}\times0.25^{\asec}$, with a position angle of $-84^{\circ}$. A bright point-like source is detected at the center of the image, with a peak flux of $47~\mu$Jy/beam and a signal-to-noise ratio of $\sim 7$. We used the \texttt{imfit} task in CASA to fit the central source with a 2D Gaussian and found the source had a major-axis FWHM of $0.33\pm0.05^{\asec}$ and a minor-axis FWHM of $0.25\pm0.03^{\asec}$, with a position angle of $39^{\circ}$. The apparent source size is thus consistent with the result of a point source convolved with the synthesized beam. The total flux measured from a 2D Gaussian fit is $53\pm10~\mu$Jy.  The final calibrated image is shown in Figure~\ref{fig:ALMA-image}.

The source appears to be slightly elongated in the North-South direction compared to the synthesized beam. To test whether this apparent elongation was a result of a binary system, we fit the source with two models: a single point source model and a binary model.  The residuals after subtracting either model are very similar, and both were within 1$\sigma$ of the distribution of noise measured in the background regions of the synthesized image. We conclude that the observations are consistent with a single point source.

%%%%%% Stephanie's Section
The combination of optical and thermal measurements
allows for constraints on the object's physical properties, under the
assumption of a thermal model with temperature distribution $T(\theta,
\phi)$, where $\theta$ and $\phi$ are the planetographic coordinates
on the object. The physical properties of primary interest are the
object's size and albedo, which can be uniquely determined
 by simultaneously solving the following equations:
\begin{eqnarray}
  \label{eqn:visibleflux}
  F_V=F_{\odot,V}{p_V\over4}\left({D\over\Delta}\right)^2\left({r\over1~\mathrm{AU}}\right)^{-2},  \\
  \label{eqn:thermalflux}
  F_{\lambda}=\int{\epsilon_{\lambda}B_{\lambda}(T(\theta,\phi))d\Omega},
\end{eqnarray}
where $F_V$, $F_{\odot,V}$, $F_{\lambda}$ are the visible, solar
visible, and thermal flux densities; $D$ is the diameter; $p_V$ is the
geometric visible albedo; $r$ and $\Delta$ are the heliocentric and
geocentric distances, respectively; $\epsilon_{\lambda}$ is the spectral
emissivity; $B_{\lambda}$ is the Planck function; and $d\Omega$ is the
solid angle subtended by the elements $d\theta$ and d$\phi$ as
seen from Earth. The form of $T=T(\theta,\phi)$ depends on the surface geography, spin rate, bolometric emissivity ($\epsilon$), thermal inertia, and shape of the object, none of which is generally
known. Therefore it is common practice to assume a simplified thermal
model.

The Standard Thermal Model (STM; \citealt[and references therein]{Lebofsky1989})
describes a spherical, non-rotating body observed at zero
phase angle and represents the hottest possible temperature
distribution. In this model, the temperature depends only on the
angular distance $\theta$ from the subsolar point where $T(\theta) =
T_{SS}\cos^{1/4}\theta$ with $T_{SS}=[(1-A)S_{\odot}/({\epsilon\eta\sigma}r^2)]^{1/4}$ and $T=0$ on the nightside. Here $A=qp_V$ is
the Bond albedo and $q$ is the phase integral. $S_{\odot}$ is the solar constant and $\eta$
is the ``beaming factor.'' $\eta$ was originally introduced to account
for surface roughness and variations in thermal inertia but also
serves to interpolate between the STM and its ``cold'' counterpart,
the Fast Rotator Model (FRM), by scaling the subsolar temperature. In
particular, $\eta<1$ results in a higher temperature than predicted
by the model while $\eta>1$ results in a lower predicted
temperature.

Obtaining several measurements spanning both sides of the peak of the
blackbody emission spectrum would allow us to leave $\eta$ as a free
parameter to fit in our model, significantly constraining the
temperature distribution on the surface. However this is not possible
with a single-wavelength measurement, and we must allow $\eta$ to explore 
its full range of 0.6 to 2.6 \citep{Mommert2012}. For the remainder of this work we use the STM as our base thermal model, but we allow $\eta$ to take any value within its allowed range. Equation (\ref{eqn:thermalflux}) then becomes:
\begin{eqnarray}
  F_{\lambda} = \frac{\epsilon_{\lambda} D^2}{2 \Delta^2} \int_0^{\pi/2}
  B_{\lambda}(T(\theta)) \sin\theta \cos\theta d\theta.
\end{eqnarray}

% For the remainder of this work we
% adopt the ``TNO-tuned'' model, equivalent to the STM with a constant
% $\eta=1.20\pm0.35$ as determined by \citet{Stansberry2008} from
% Spitzer observations of several Centaurs and TNOs. Under this
% model Equation (\ref{eqn:thermalflux}) becomes:
% \begin{eqnarray}
%   F_{\lambda} = \frac{\epsilon D^2}{2 \Delta^2} \int_0^{\pi/2}
%   B_{\lambda}(T(\theta)) \sin\theta \cos\theta d\theta.
% \end{eqnarray}

Adopting the STM requires some assumptions
regarding the nature of the object's thermal emissions. First, we
assume a bolometric emissivity $\epsilon=0.9\pm0.1$, a typical assumption for TNO thermal models. While the thermal emissivity can be treated as constant at wavelengths \mbox{$\lesssim 350\mu$m}, 
\citet{Fornasier2013} showed that the thermal emissivity is suppressed at longer wavelengths, such as the 1.3mm wavelength corresponding to ALMA's 233 GHz band used in our measurement. 
We adopt a value of $\epsilon_{\lambda}=0.68$ at 1.3mm, the average value measured by \citet{BrownButler2017} with ALMA at 233 GHz for four TNOs with sizes comparable to \UZ. We keep $\epsilon=0.9\pm0.1$ as our bolometric emissivity in the surface temperature distribution.
Second, we assume that the phase integral
$q=0.8$ as derived in \citet{Stansberry2008} for large, bright TNOs.
Varying $q$ from $0.4$ to $0.8$ results in $<1\%$
variation in albedo for low-albedo TNOs.  The phase angle for an object
at $\sim$92 AU never exceeds $1^{\circ}$; thus we may neglect any
effects arising from a changing phase angle and set the phase angle equal to zero.

Estimation of the uncertainties in the calculated diameter and albedo
were performed following the procedure outlined in
\citet{Mommert2012}. We employed a Monte Carlo simulation using
1000 clones, where each clone was generated by varying the observed
flux densities at both the thermal and optical wavelengths, the
heliocentric and geocentric distances associated with the optical
measurements, the bolometric emissivity $\epsilon$, and the beaming factor $\eta$. Our uncertainties are dominated by
the statistical uncertainty in our flux measurements. Each parameter, with the exception of $\eta$, was varied
randomly according to a normal distribution defined by its nominal
value and 1$\sigma$ uncertainty. $\eta$ was varied according to a uniform distribution from 0.6 to 2.6. The uncertainties in the diameter and albedo were
then defined by the lower and upper values that included 68.2\% of the
clones, centered on the peaks of the resulting distributions of the
two parameters. 

Under these assumptions we measure the geometric albedo and diameter of \UZ\ to be

\begin{eqnarray*}
p_{V} & = & 13.1^{+3.3}_{-2.4}\mathrm{(stat)}^{+2.0}_{-1.4}\mathrm{(sys)}\% \\
D & = & 635^{+57}_{-61}\mathrm{(stat)}^{+32}_{-39}\mathrm{(sys)}~\mathrm{km}.
\end{eqnarray*}
Here the quoted statistical uncertainty is due to the uncertainties in both the visual and thermal flux measurements, as well as uncertainties in the helio- and geo-centric distance measurements. The quoted systematic uncertainty is due to variation of the model parameters $\eta$ and $\epsilon$.

\begin{figure}[htbp]
\begin{center}
\includegraphics[width=\columnwidth]{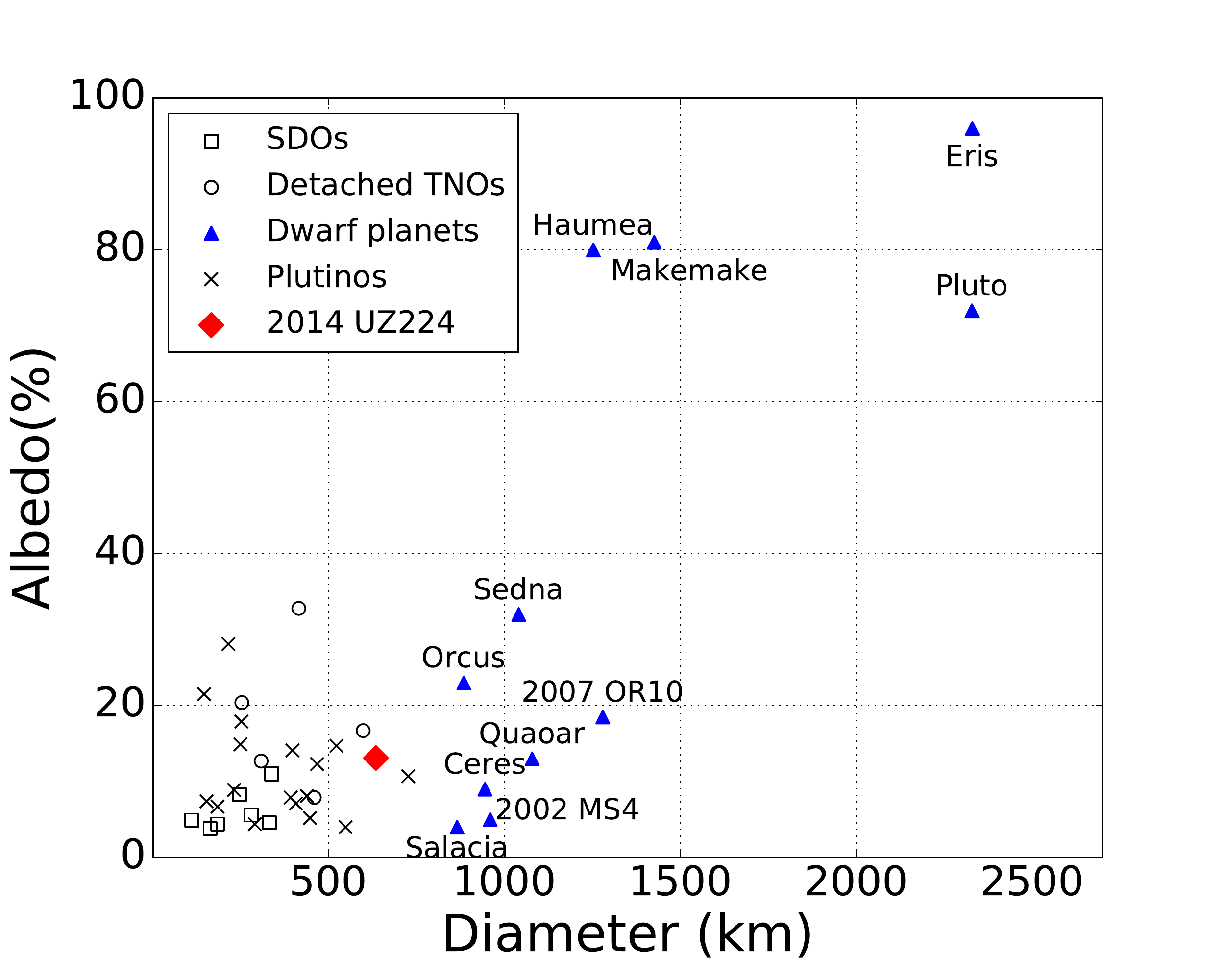}
\caption{\label{fig:size_albedo} Size-albedo relations for selected TNO populations including detached and scattered disk objects \citep{Li2006, SantosSanz2012}, Plutinos \citep{Mommert2012}, and dwarf planets 
\citep{Sicardy2011, Fornasier2013, Pal2012, Buratti2016, BrownButler2017}. The object \UZ\ is shown as a filled diamond.}
\end{center}
\end{figure}

As shown in Fig.~\ref{fig:size_albedo}, the measured albedo is higher than that of rocky bodies such as asteroids, and of typical classical KBOs, 
yet notably smaller than ice-rich dwarf planets Eris (96\%, \citet{Sicardy2011}), Haumea (80\%, \citet{Fornasier2013}), Pluto (72\%, \citet{Buratti2016}) and Sedna (32\%, \citet{Pal2012}), suggesting that
\UZ\ has a mixed ice-rock composition. An object of this composition and size is likely to have enough self-gravity to reach an approximately spherical shape in hydrostatic equilibrium \citep{Tancredi2008}, 
making \UZ\ a candidate dwarf planet. 

%%%%%%%%%%%%%%%% end of Stephanie's text

\section{Conclusions}
\label{sec:conclusion}

We have reported the discovery of \UZ\ (``DeeDee'') a trans-Neptunian
object discovered at 92 AU from the Sun. This object has
an estimated size $D =635\pm 70$ km and albedo $p_V=13\pm 4 \%$, and is
most likely a dwarf planet with a mixed ice-rock composition. This
discovery adds to the growing inventory of dwarf planets in the
outer Solar System, and indicates that the TNO population displays a
nearly continuous distribution of size and albedo.

Neither the orbital nor the physical properties of \UZ\ are surprising, as they are in the range of other well-characterized detached TNOs discovered closer to the Sun.  The population of detected TNOs is of course strongly biased toward those that are large, near perihelion, and/or have high albedo.  Current surveys such as the DES now have the depth and area coverage to discover the counterparts of known objects that are well beyond perihelion.  It is also noteworthy that the ALMA facility is easily capable of radiometric detection of a 600~km body at $>90$~AU distance. Hence it will be possible to establish sizes and albedos for nearly every body detectable in the visible by DES and similar surveys.  As these surveys progress, we will be able for example to determine whether the very high albedo of Eris is characteristic of large bodies at this distance, or whether flux selection has led to the first discovery being atypical.

\section{Acknowledgements}
\label{sec:acknowledgements}

D.W.G. and F.C.A. are partially supported by NSF grant AST-1515015. G.M.B. and M.S. are partially supported by NSF grant AST-1515804.
S.J.H. and J.C.B. are supported by NSF-GRFP grant DGE-1256260. This work used the Extreme Science and Engineering Discovery Environment (NSF grant number ACI-1053575).

Funding for the DES Projects has been provided by the DOE and NSF(USA), MEC/MICINN/ MINECO (Spain), STFC (UK), HEFCE (UK). NCSA (UIUC), KICP (U. Chicago), CCAPP (Ohio State), 
MIFPA (Texas A\&M), CNPQ, FAPERJ, FINEP (Brazil), DFG (Germany) and the Collaborating Institutions in the Dark Energy Survey.

The Collaborating Institutions are Argonne Lab, UC Santa Cruz, University of Cambridge, CIEMAT-Madrid, University of Chicago, University College London, 
DES-Brazil Consortium, University of Edinburgh, ETH Z{\"u}rich, Fermilab, University of Illinois, ICE (IEEC-CSIC), IFAE Barcelona, Lawrence Berkeley Lab, 
LMU M{\"u}nchen and the associated Excellence Cluster Universe, University of Michigan, NOAO, University of Nottingham, Ohio State University, University of 
Pennsylvania, University of Portsmouth, SLAC National Lab, Stanford University, University of Sussex, Texas A\&M University, and the OzDES Membership Consortium.

The DES Data Management System is supported by the NSF under Grant Number AST-1138766. The DES participants from Spanish institutions are partially 
supported by MINECO under grants AYA2012-39559, ESP2013-48274, FPA2013-47986, and Centro de Excelencia Severo Ochoa SEV-2012-0234. Research leading 
to these results has received funding from the ERC under the EU's 7$^{\rm th}$ Framework Programme including grants ERC 240672, 291329 and 306478.

This paper makes use of the following ALMA data: ADS/JAO.ALMA\#2015.A.00023.S. ALMA is a partnership of ESO (representing its member states),
NSF (USA) and NINS (Japan), together with NRC (Canada), NSC and ASIAA (Taiwan), and KASI
(Republic of Korea), in cooperation with the Republic of Chile. The Joint ALMA Observatory is operated by ESO, AUI/NRAO and NAOJ. The National Radio Astronomy Observatory is a facility of the National
Science Foundation operated under cooperative agreement by Associated Universities, Inc.

\software{Astropy \citep{astropy},
          Matplotlib \citep{matplotlib},
          Numpy \citep{numpy},
          Pandas \citep{pandas}}
 
%\bibliography{apj-jour,trojans}
  \newcommand{\noop}[1]{}

 \end{document}